\documentstyle[12pt,aasms4,graphicx]{article}
\lefthead{Baella}
\righthead{StH$\alpha$63: a new yellow symbiotic star}

\slugcomment{Submitted to the Astronomical Journal}

\begin{document}

\title{Searching for new yellow symbiotic stars: positive identification of
  StH$\alpha$63}

\author{N.O. Baella$^{1,2}$, C.B. Pereira$^{3}$, L.F. Miranda$^{4}$, and A. Alvarez-Candal$^{3}$} 
\affil{1: Unidad de Astronom\'{\i}a, Instituto Geof\'{\i}sico del Per\'u, Lima, Per\'u\\
2: Departamento de Ciencias, Secci\'on F\'{\i}sica, Pontificia Universidad
Cat\'olica del Per\'u, Apartado 1761, Lima, Per\'u\\
3: Observat\'orio Nacional/MCTI, Rua Gen. Jos\'e Cristino, 77,  20921-400,
Rio de Janeiro, Brazil\\
4: Instituto de Astrof\'{\i}sica de Andaluc\'{\i}a\,-\,CSIC, C/ Glorieta de la Astronom\'{\i}a
s/n, E-18008 Granada, Spain\\
Email:nobar.baella@gmail.com,claudio@on.br,lfm@iaa.es,alvarez@on.br}

\bigskip

\author{Submitted to The Astronomical Journal}

\altaffiltext{1} {Based on the observations collected at the Centro
  Astron\'omico Hispano-Alem\'an, Calar Alto, jointly operated by the
  Max-Planck-Institut f\"ur Astronomie (Heidelberg) and the Instituto de
  Astrof\'{\i}sica de Andaluc\'{\i}a (CSIC), and at the 4.1m telescope at
  Cerro Pach\'on Observatory, Chile.}

\begin{abstract}

\par Yellow symbiotic stars are useful targets to probe whether mass
transfer has happened in these binary systems. However, the number of
known yellow symbiotic stars is very scarce. We report spectroscopic
observations of five candidate yellow symbiotic stars selected by
their position in the 2MASS (J$-$H) {\sl versus} (H$-$K$_{s}$) diagram
and included in some emission-line catalogs. Among the five
candidates, only StH$\alpha$63 is identified as a new yellow symbiotic
star because of its spectrum and its position in the
[TiO]$_1$\,--\,[TiO]$_2$ that indicates a K4--K6 spectral type. In
addition, the derived electron density ($\sim$ 10$^{8.4}$\,cm$^{-3}$)
and several emission line intensity ratios provide further support for
that classification. The other four candidates are rejected as
symbiotic stars because three of them actually do not show emission
lines and the fourth one shows only Balmer emission lines. We also found that the WISE
W3--W4 index clearly separates normal K-giants from yellow symbiotic
stars and, therefore, can be used as an additional tool to select
candidate yellow symbiotic stars.

\end{abstract}


\section{Introduction}

\par Symbiotic stars are interacting binary systems formed by a red
giant star and a hot source (in most cases a white dwarf) that ionizes
the stellar wind from the cool component. They can be divided in three
different infrared types, D, D$^{'}$ and S, that are related to the
nature of the cool giant in the system, orbital separation and
physical conditions in the emitting nebulae (see Mikolajewska
1997). The number of currently identified symbiotics is much smaller
than the predicted one. The catalog of Belczy\'nski et al. (2000)
contains 173 Galactic symbiotics, while the predicted number ranges
from 3$\times$10$^{3}$ to 4$\times$10$^{5}$ (Allen 1984; Kenyon et
al. 1993; Munari \& Renzini 1992; Magrini et al. 2003). Recently,
several surveys have discovered new symbiotic stars increasing their
number. Corradi et al. (2008, 2010) and Rodr\'{\i}guez-Flores et
al. (2014) discovered 12 S-type and 4 D-type new symbiotics by
combining the IPHAS survey and the position of the candidate stars in
the 2MASS color-color diagram. Miszalski et al. (2013) used the
SuperCosmos H$\alpha$ Survey (SHS) (Parker et al. 2005), selected
H$\alpha$ emission candidates towards the Galactic bulge, and
discovered 13 S-type, 6 D-type, and 1 D$^{'}$-type new
symbiotics. Miszalski \& Mikolajewska (2014), also based on the SHS
catalogue, found 11 S-type and 1 D-type new symbiotics in the southern
hemisphere. Therefore, a total of 48 new symbiotic stars have been
added to the 173 previously known ones.

\par Among S-type symbiotic stars, there is a group known as ``yellow
type symbiotics'' in which the cool component is a K-type star rather
than an M-type giant. Yellow symbiotics are useful objects to probe
whether mass transfer has taken place in the past of these binary
systems. The atmosphere of the K-type component of yellow symbiotics
will be free from the strong molecular opacities due to TiO, CN and
C$_{2}$ absorption features that are present in the S-type symbiotics
with an M-type giant, thereby enabling the determination of stellar
abundances based on atomic lines. Studies of chemical abundances in
yellow symbiotics indeed show that these stars are enriched in
s-process elements. Because these stars are not luminous enough to be
self-enriched in s-process elements, the observed enrichment can be
attributed to mass transfer from the former AGB star (now the white
dwarf) of the system (see Smith et al 1996, 1997; Pereira \& Roig
2009). Therefore, yellow symbiotics are valuable objects to test
models of AGB nucleosynthesis (Busso et al. 2001). However, they are
very rare objects: only 12 Galactic yellow symbiotics are included in
the catalogue of Belczy\'nski et al. (2000) that represent only $\sim$
5.4\% of the total number of 221 S-type symbiotics identified to date
(see above). Moreover, yellow symbiotics are members of the field
Galactic halo population (Schmid \& Nussbaumer 1993; Jorissen
2003). Therefore, there is not a specific Galactic direction where
they can be surveyed and, in consequence, are very difficult to
find. In fact, among the $\sim$ 3.6 million stellar spectra obtained
with the LAMOST survey, only one new yellow symbiotic was found:
LAMOST J12280490-014825.7, a K3 star with a high radial velocity and
at high Galactic latitude (Li et al. 2015), similarly to some yellow
symbiotics previously analyzed (Pereira \& Roig 2009).

\par Due to the rarity of these objects and absence of a particular
Galactic location, surveys to search for new yellow symbiotics were
inhibited in the past. However, Baella (2012) showed that combining
their infrared properties and emission spectrum, yellow symbiotics are
located in a particular region of the 2MASS color-color diagram. This
led to the discovery of SS\,383 as our first candidate yellow
symbiotic that has been identified following the criteria above
mentioned (Baella et al. 2013, hereafter BPM13).

\par In this work we report further spectroscopic results of our
survey specifically dedicated to search and to identify new yellow
symbiotics. Our major result is the identification of StH$\alpha$63 as
a new yellow symbiotic star.

\section{Target selection}

\par The selection of the candidates followed the same method as we
applied for the identification of SS\,383 as a new candidate yellow
symbiotic star (BPM13). Basically our method rely on the two observed
properties of the yellow symbiotic stars, the emission-line spectrum
and the K-type continuum. It was already well stablished that
symbiotics occupy particular regions in the 2MASS $(J-H)$ {\sl versus}
$(H-K_{s})$ diagram (Phillips 2007; Corradi et al. 2008,
2010). Furthermore, Baella (2012) showed that, after reddening
correction, yellow symbiotics are clustered around $(H-K_{s})$ $\sim$
0.2 and $(J-H)$ $\sim$ 0.8 because their positions in the two color
diagram are mainly determined by the K-type component and they are
clearly separated from S-type symbiotics with M-type components. By
using the emission line catalogs of Stephenson \& Sanduleak (1977),
Stephenson (1986), Schwartz, Persson \& Hamamm (1990) and Kohoutek \&
Wehmeyer (1997) we have identified five more stars that are located in
the region of yellow symbiotics of the 2MASS color-color diagram and,
therefore, are good candidates to be yellow symbiotics. Table 1 lists
the candidates and Figure 1 show their position in the $(J-H)$ {\sl
  versus} $(H-K_{s})$ diagram. To determine the corresponding reddening value for each
object, we used the Galactic Dust Reddening and Extinction Service of IRSA
(Infrared Science Archive): http://irsa.ipac.caltech.edu/applications/DUST/ to
obtain the ``E(B-V) Reddening'' values, and converted E(B-V) to A(J), A(H), and
A(Ks) extinctions using the relationships given by Bilir et al. (2008), which
are used to draw the corresponding reddening vectors in Figure\,1. In
particular, for StH$\alpha$63 we used a mean E(B-V)\,=\,0.14 (Schlafly \& Finkbeiner
2011; Schlegel et al. 1998), and obtained A(J-H)\,=\,0.05
and A(H-Ks)\,=\,0.03.

\section{Observations \& Reductions}

\par Spectroscopic observations of the five targets in Table 1 were
performed on 2013 and 2014 with the Cassegrain Twin Spectrograph
(TWIN) attached to the 3.5\,m telescope of the Calar Alto Observatory
(Spain), and with the Goodman spectrograph attached to the 4.1\,m
telescope of the SOAR Observatory (Chile). Table 1 also includes the
date of observations and the used spectrograph. TWIN includes two
separate spectroscopic channels (blue and red) behind the common
entrance slit aperture. The detectors were a SITe CCD 22b in the blue
channel and a SITe CCD 20b in the red channel. We used gratings T08
(blue) and T04 (red) to cover the spectral ranges 3200--5800\,\AA\,
and 5500--7600\,\AA\,, respectively, at a spectral resolution of
1.08\,\AA\,pixel$^{-1}$ in both channels. The slit was oriented
east--west and its width was 2$''$.  The Goodman spectrograph was used
with the Farchild CCD 486 and three gratings: blue, mid, and red mode,
covering the spectral ranges 3550--6300\,\AA\,, 4500--7250\,\AA\,, and
6400--9150\,\AA\,. respectively, at a spectral resolution of
0.65\,\AA\,pixel$^{-1}$ in the three spectral ranges. The slit was
oriented north--south and its width was 1$''$.  Spectra of each target
were obtained with several exposure times between 15 and 1200 seconds,
the short exposures to avoid saturation of possibly strong emission
lines (e.g., H$\alpha$), the long exposures to detected faint emission
lines.

\par The spectra were reduced using standard IRAF tasks, from bias
subtraction and flat-field correction, through spectral extraction and
wavelength and flux calibration. Spectrophotometric standards from
Massey et al. (1988) and Hamuy et al. (1994) were also observed with
the same instrumental configuration before and after the objects for
flux calibration.

\section{StH$\alpha$63 : a new yellow symbiotic star}

\par Figure 2 presents the spectrum of StH$\alpha$63 and Table 2 lists
the observed fluxes of the identified emission lines. The spectrum of
StH$\alpha$63 shows recombination Balmer, He\,{\sc i}, and He\,{\sc
  ii}\,$\lambda$4686 emission lines, and forbidden [O\,{\sc
    iii}]\,$\lambda$4363,4959,5007 emission lines. The absorption
spectrum presents the TiO bands at 6200, 7125, and 7160\AA , and
absorption lines due to Fe\,{\sc i} at 4202,5227,5269,5227,5415\,\AA\,,
Ca\,{\sc i} at 4226\,\AA\,, Mg\,{\sc i} at 5183\,\AA\,, Cr\,{\sc i} at
5206\,\AA\,, Na\,{\sc i} at 5890\,\AA\,, and a strong absorption around
6500\,\AA\, that is due to contributions of several strong absorption
lines of Fe\,{\sc i} at 6494,6496\,\AA\,, Ba\,{\sc ii} at 6497\,\AA\,, and
Ca\,{\sc i} at 6499\,\AA. The spectrum clearly points to a symbiotic
nature for StH$\alpha$63. Moreover, the spectrum is very similar to
that of the yellow symbiotic CD-43$^\circ$14304 with a K7 spectral
type (Schmid \& Nussbaumer 1993; Pereira et al. 1999; M\"urset \&
Schmid 1999), and to that of the candidate yellow symbiotic SS\,383
with a spectral type K7--M0 (BPM13). These similarities strongly
suggest a K spectral type for StH$\alpha$63 and, in consequence, that
StH$\alpha$63 is a yellow symbiotic.

\par To obtain a more precise estimate for the spectral type of
StH$\alpha$63 we proceeded as we did for SS\,383: we measure the
quantitative TiO indexes (Kenyon \& Fern\'andez-Castro 1987) and put
StH$\alpha$63 in the [TiO]$_{1}$\,--\,[TiO]$_{2}$ diagram. Figure 3
shows the position of StH$\alpha$63 in this diagram, which lies
between those of AG Dra and TV CVn, two stars with spectral types
K4\,III and K5.3\,III, respectively (Kenyon \& Fern\'andez-Castro
1987). Therefore, we estimate a spectral type between K4 and K6 for
StH$\alpha$63 that confirms the yellow symbiotic nature of the object.

\par Some physical parameters of StH$\alpha$63, such as the color
excess $E(B-V)$, optical depth in H$\alpha$ ($\tau_{H\alpha}$),
electron density ($N_{e}$), and infrared type based on He\,{\sc i}
line intensity ratios can be obtained directly from the observed
spectrum. To obtain these parameters we follow the same basic
procedures as outlined in BPM13 (see also Guti\'errez-Moreno \& Moreno
1996, and Proga et al. 1994). Table 3 shows the results. The electron
density derived from the [O\,{\sc iii}] emission lines intensity ratio
is $log N_{e}$ $\sim$ 8.4 (see Table 3), a value that is typical of
S-type symbiotics ($log N_{e}$ $\sim$ 8--10) but larger than the values
found in D-type symbiotics ($log N_{e}$ $\sim$ 6--7, e.g., Schmid \&
Schild 1990; Mikolajeska \& Kenyon 1992). As for the infrared (D or S)
type (Proga et al. 1994), the line ratios $I(He\,{\sc
  i}\lambda6678)/(He\,{\sc i}\lambda5876)$ and $I(He\,{\sc
  i}\lambda7065)/(He\,{\sc i}\lambda5876)$ corrected for reddening are
0.31 and 0.78, respectively. The value of the $I(He\,{\sc
  i}\lambda6678)/(He\,{\sc i}\lambda5876)$ ratio is not typical for
S-type symbiotics according to Proga et al. (1994), although PU Vul
investigated in their sample also presents a similar ratio of
0.30. The $I(He\,{\sc i}\lambda7065)/(He\,{\sc i}\lambda5876)$
intensity ratio in StH$\alpha$63 is similar to that observed in the
yellow symbiotic AG Dra (Proga et al. 1994).

\par Finally, our spectrum allow us to analyze the position of
StH$\alpha$63 in the [O\,{\sc iii}]5007/H$\beta$ {\sl versus} [O\,{\sc
    iii}]4363/H$\gamma$ diagnostic diagram that was previously used in
our studies of planetary nebulae (Pereira \& Miranda 2005; Pereira et
al. 2010) and of the yellow symbiotic SS\,383 (BPM13). The line
intensity ratios (see Table 2) place StH$\alpha$63 in the region of
the diagram occupied by S-type symbiotics (see BPM13, their Figure\,5
) thus further reinforcing our conclusion of the symbiotic nature for
this object.

\section{The spectra of [KW97] 37-26, [KW97] 61-27, StH$\alpha$116 and  SS\,360.}

\par Figure 4 presents the spectra of the four candidates in Table 1
that were rejected as symbiotics either due to the absence of emission
lines or, when present, because they were not the appropriate ones to
classify the star as a symbiotic.  Noticeably, [KW97] 37-26, [KW97]
61-27, and StH$\alpha$116 do not show the H$\alpha$ emission line even
though the are included in H$\alpha$ emission-line star catalogs (see
MacConnel 1981; Stephenson 1986; Kohoutek \& Wehmeyer 1999). In the
cases of [KW97] 61-27 and StH$\alpha$116 we note that these two stars
are characterized by very strong TiO absorption bands (Figure 4) that
produce an ``apparent emission peak'' around 6540\AA\,, bluewards the
position of H$\alpha$, which may be confused with the H$\alpha$
emission line in very low resolution (objective-prism) spectra. Only SS\,360, discovered
as an emission-line star by Stephenson \& Sanduleak (1977), presents
Balmer emission lines. However, the other typical emission lines of
symbiotics (e.g., [O\,{\sc iii}]) are absent, discarding a symbiotic
classification.

\par The spectrum of [KW97] 37-26 (=\,[M81]
I-734\,=\,BD+00$^\circ$4203) is very similar to those of HD\,1069 and
HD\,26946 with K2\,I and K3\,III spectral types, respectively (Jacoby
et al. 1984), if we apply to these two stars a reddening of 1.4 as
used in Figure 1 for [KW97] 37-26. This points out that [KW97] 37-26
is an early K supergiant or giant star. A more precise classification
is not possible because our spectra do not cover the oxygen line at
7774\AA\,, the calcium infrared triplet and/or the CN band at
7925\AA\, that are useful spectral features to discriminate among
supergiants, giants and dwarf stars (see Pereira \& Miranda 2007). The
spectra of StH$\alpha$116 (=\,BD+14$^\circ$2953) and [KW97] 61-27 are
similar to that of SAO 21753, a K7\,III star, whereas the spectrum of
SS\,360 is similarities to that of the M3\,III star SAO\,63340 (see
Jacoby et al. 1984 for the spectra of the used comparison stars).

\section{The W34-index from WISE bands as an additional constraint to search
for candidates to S-type yellow symbiotic stars.}

\par As shown above, SS\,360 fullfils the two criteria used by us to
select candidate yellow symbiotics but its spectrum rules out this
classification. This suggests to look for additional constraints for a
better selection of candidates. We have explored the data provided in
the WISE (Wide-field Infrared Survey Explorer) archive. The WISE
satellite scanned the whole sky in the 3.4, 4.6, 12, and 22$\mu$m
bands (Wright et al. 2010) commonly named W1, W2, W3 and W4,
respectively. We used the W3--W4 (W34) index and created five
histograms to analyze the possibility of distinguishing true S-type
yellow symbiotics stars among the candidate objects in Figure 1.

\par Figure 5 shows these five histograms. Histogram A presents a
sample of about 340 K-giant stars with effective temperature between
3700 and 5200\,K (Hekker \& Mel\'endez 2007). It shows that the
probability of finding K-giant stars with W34 $>$ 0.4 is very
low. Histogram B presents the distribution of S-type symbiotic stars
from Belczynski et al. (2000). These objects cover the entire range of
W34 values in the figure, although there is a small increase of the
number of S-type symbiotics in the W34 0.4--0.8 range. This could be
an indication that S-type symbiotics have an emission excess between
12$\mu$m and 22$\mu$m. To check this possibility, histogram C shows a
sub-sample of objects of histogram B, namely, those S-type symbiotics
that contain giant stars with effective temperatures higher than
3500\,K; these symbiotics present W34 values higher than 0.4. It is
worth noting that effective temperatures higher than $\sim$ 3500\,K
are required for yellow symbiotics. Histogram D shows a sample of nine
true S-type yellow symbiotics. Remarkably, they all have W34 $>$ 0.4,
indicating that the difference between S-type yellow symbiotics and
``normal'' K-giants (without symbiotic activity) is the W34
value. This is probably due to the presence of a dust shell in true
S-type yellow symbiotics. The dust shell could add an excess of
emission between 12$\mu$m and 22$\mu$m, that is revealed through the
W34 index that appears as an useful tool to distinguish between yellow
symbiotic and ``normal'' K-giants.

\par To investigate these results further, we construct the histogram E with
the five objects of Table 1 and SS\,383 (BPM13). Only SS\,383 and
StH$\alpha$63 have W34 $>$ 0.4, which is compatible with the yellow symbiotic
classification for StH$\alpha$63, and provides further support for a yellow
symbiotic nature of the candidate SS\,383. Moreover, SS\,360 has W34 $\sim$
0.21 and, therefore, it is not located in the region of yellow symbiotics, in
agreement with its non-yellow symbiotic nature stablished above, even though
it is an emission-line star and its position on the 2MASS two color diagram
favors such a classification. From these findings, we conclude that the W34
index can be used as a useful tool to identify candidate yellow symbiotics, in
combination with emission-line star catalogs and the 2MASS color-color diagram.

\section{Conclusions}

\par In this work we presented spectroscopic observations of five
emission-line stars that are candidates for yellow symbiotics on the
basis of their position in the 2MASS color-color diagram. Only one of
them, StH$\alpha$63, has been positively identified as a new yellow
symbiotic. Its spectrum and position in the
[TiO]$_{1}$\,-\,[TiO]$_{2}$ diagram indicates a spectral type between
K4 and K6. In addition, He\,{\sc i} emission line ratios, derived
electron density, and position in the [O\,{\sc iii}]5007/H$\beta$ {\sl
  versus} [O\,{\sc iii}]4363/H$\gamma$ diagram provide further support
for the yellow symbiotic nature of StH$\alpha$63. The other four
candidates are rejected as symbiotics because three of them do not
show emission lines (although they were classified as emission-line
stars) or one shows only Balmer emission lines.

\par Our survey for yellow symbiotics based on their particular
position in the 2MASS color-color diagram and four emission-line star
catalogs raised only six candidates (SS\,383 and the five stars
analyzed in this work). This small number is not surprising given the
rarity of yellow symbiotics. Two of them (SS\,383 and StH$\alpha$63)
have been identified as new yellow symbiotics, which provides strong
support to our method of target selection. The success of this survey
becomes more clear if one takes into account that three of the six
selected objects were erroneously classified as emission-line stars.

\par We used the WISE archive to explore the possibility of distinguishing yellow
symbiotic stars. We found that yellow symbiotics have a W3--W4 (W34) index $>$ 0.4 and
are clearly separated from ``normal'' K-giant stars. Therefore, the W34 index
is an useful tool to improve the selection of candidate yellow symbiotics.

\acknowledgments

\par We are very grateful to the staff at the Calar Alto Observatory
by the excellent observations, and to David Sanmartin (SOAR) for the
excellent support given during the observation of SS\,360. LFM acknowledges partial
support by grant AYA2011-30228-C03.01 and AYA2014-57369-C3-3-P of the Spanish
MINECO, both co-funded by FEDER funds.

\begin{table}
\caption{List of objects spectroscopically observed, with the equatorial coordinates,
 V-magnitudes, used spectrograph, and observed infrared color-indexes.}
\begin{tabular}{|l|c|c|c|c|c|c|c|}\hline

Star & Date Obs & $\alpha_{\sl 2000}$  & $\delta_{\sl 2000}$  & V\,(GSC)$^{a}$ & T/G$^{c}$ & J$-$H$^{d}$ 
& H$-$K$^{d}$ \\\hline 
%                  &             &            &               &       &   &      &      \\\hline
$[$KW97$]$ 37-26  & 2013 Jun 26 & 19 26 12.5 & $+$00 55 20.3 &  10.0 & T & 0.88 & 0.24 \\ 
$[$KW97$]$ 61-27  & 2013 Jun 26 & 22 07 38   & $+$49 00 18.0 &  11.8 & T & 0.92 & 0.26 \\
SS\,360           & 2014 Jul 18 & 18 17 37.7 & $-$28 37 05.6 &  12.3 & G & 0.93 & 0.25 \\
StH$\alpha$63     & 2013 Oct 15 & 07 58 05.9 & $-$07 43 55.5 &  13.2$^{b}$ & T & 0.80 & 0.25\\
%StH$\alpha$78    & 2014 May 04 & 09 59 04.3 & $+$07 22 47.9 &  13.4 & T & 0.64 & 0.20 \\
StH$\alpha$116    & 2013 Jun 26 & 15 52 19.7 & $+$14 15 19.8 &  10.4 & T & 0.79 & 0.22 \\\hline

\end{tabular}

a: The Guide Star Catalog, Lasker et al. (2008)\par
b: Nomad Catalog, Zacharias et al. (2005)\par 
c: T: Twin spectrograph; G: Goodmann spectrograph\par
d: Cutri et al. (2003)\par
\end{table}

\clearpage

\begin{table}
\centering
\caption{Observed emission line fluxes (relative to F(H$\beta$) = 100) in StH$\alpha$63.}
\begin{tabular}{ccc}\hline
Wavelength (\AA\ ) & Identification & F($\lambda$) \\\hline
%                   &                & (Observed)   \\\hline
3869 & [Ne\,{\sc iii}]&  7.8 \\
3889 & He\,{\sc i}    &  6.3 \\
3970 & H$\epsilon$    &  7.5 \\
4101 & H$\delta$      & 13.9 \\
4340 & H$\gamma$      & 29.6 \\
4363 & [O\,{\sc iii}] & 17.0 \\
4471 & He\,{\sc i}    &  5.0 \\
4686 & He\,{\sc ii}   &  6.6 \\
4861 & H$\beta$       & 100.0\\
4959 & [O\,{\sc iii}] &  6.6 \\
5007 & [O\,{\sc iii}] & 25.3 \\
5876 & He\,{\sc i}    & 16.2 \\
6563 & H$\alpha$      & 596.0\\
6678 & He\,{\sc i}    &  7.4 \\
7065 & He\,{\sc i}    & 20.2 \\
\hline
     &                &        \\
F(H$\beta$) erg\,cm$^{-2}$\,s$^{-1}$ & 5.1$\times$10$^{-14}$ \\ 
     &                &        \\
\hline
\end{tabular}
\end{table}

\begin{table}
\centering
\caption{Physical parameters of StH$\alpha$63.}
\begin{tabular}{ c c }\hline\hline
Parameter                   & Value        \\\hline
E(B-V)                      & 0.19$\pm$0.04 \\
A$_{V}$                      & 0.59$\pm$0.12$^{a}$ \\    
$\tau$$_{H\alpha}$            & 1.1$\pm$0.1 \\
log(N$_{e}$)$^{b}$ cm$^{-3}$  & 8.4\\
He\,{\sc i} $\lambda$6678/$\lambda$5876$^{c}$ & 0.31 \\
He\,{\sc i} $\lambda$7065/$\lambda$5876$^{c}$  & 0.78 \\
Spectral Type$^{d}$          &  K4\,-\,K6\\\hline
\end{tabular}

$^{a}$ Using R=3.1\par
$^{b}$ From the [O\,{\sc iii}] emission lines and T$_{e}$ = 10\,000\,K\par
$^{c}$ Reddening corrected\par
$^{d}$ From [TiO]$_{1}$ and [TiO]$_{2}$ indexes\par
\end{table}

{}

\clearpage

%\begin{figure*}
%   \begin{center}
%   \includegraphics[angle=0,width=180mm,clip=]{miranda_fig2_v1.eps}
%      \caption{VLA image of IRAS~18197$-$1118 at 8.46 GHz in contours and grey
%        scale (left panel), and grey scale (right panel). The contours are -4,
%        4, 6, 8, 10, 12,  15, 20, 30, 40, 50, 60, 70, 80, 90, 100,  and 110
%        times 0.079 mJy beam$^{-1}$, the rms noise of the image. The greyscale
%        values are given in the upper wedge in units of mJy\,beam$^{-1}$. The
%        synthesized beam, with dimensions of 0.51$\times$0.28\,arcsec$^2$  and
%        major axis at position angle $77^\circ$, is shown in the bottom
%        left corner of each panel.}
%    \end{center}
%   \end{figure*}

\begin{figure*}
%\begin{center}
   \includegraphics[angle=0,width=150mm,clip=]{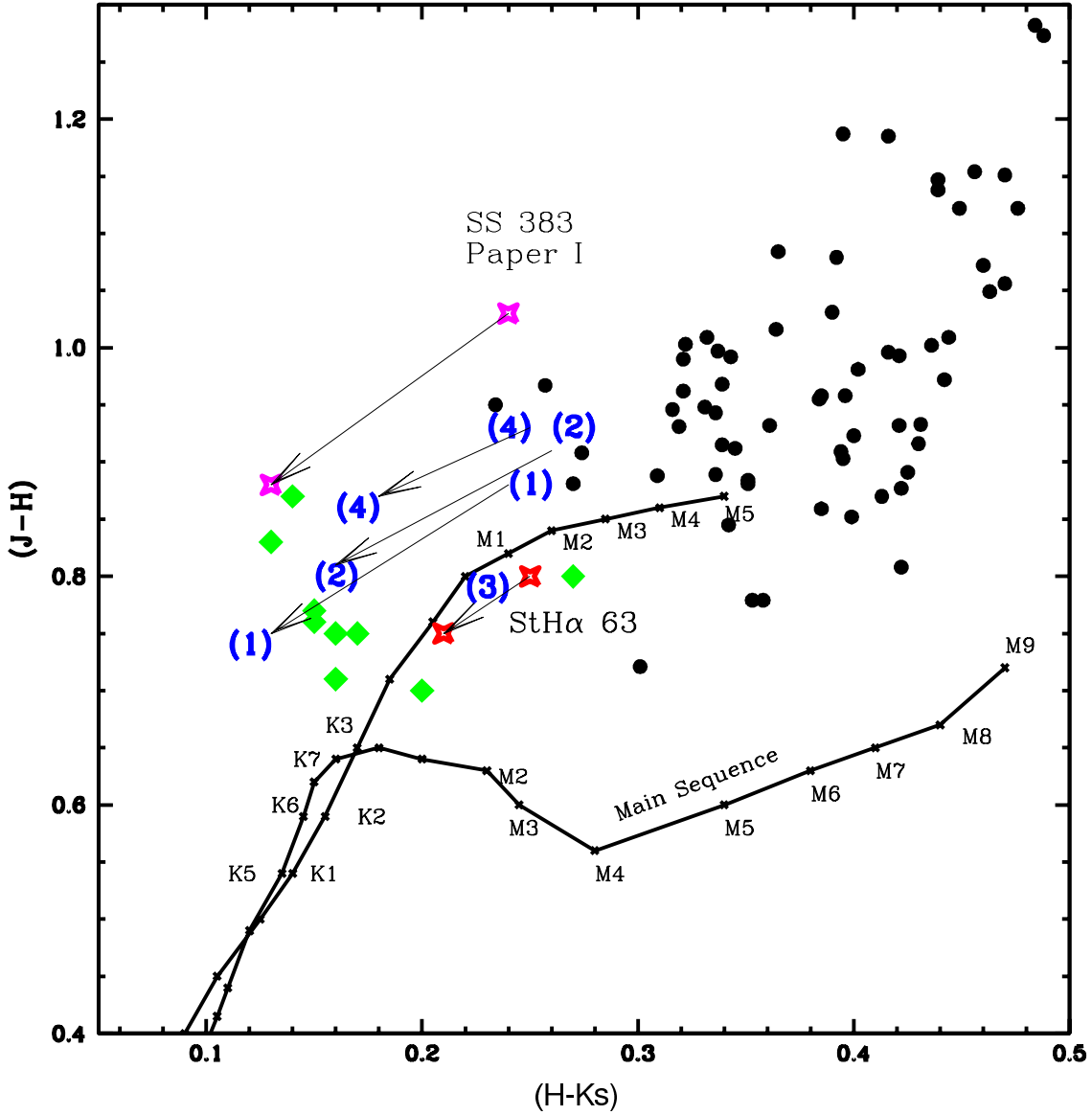}
   \caption{Position of the candidate yellow symbiotic stars [KW97] 37-26
  (1), [KW97] 61-27 (2), StH$\alpha$116 (3), SS\,360 (4), and
  StH$\alpha$63 (red star) in the 2MASS color-color diagram. The
  arrows connect the observed and reddening corrected ($J-H$) and
  ($H-K_{s}$) colors. For StH$\alpha$116 we do not show any correction
  due to the low extinction in the direction of this object. Green
  diamonds represent the reddening corrected infrared colors of seven
  S-type yellow symbiotic stars. SS\,383 (magenta star) previously
  identified as a candidate yellow symbiotic is also shown. Black
  circles represent the S-type symbiotics.}
%\end{center}
\end{figure*}

\begin{figure*}
 \includegraphics[angle=0,width=150mm,clip=]{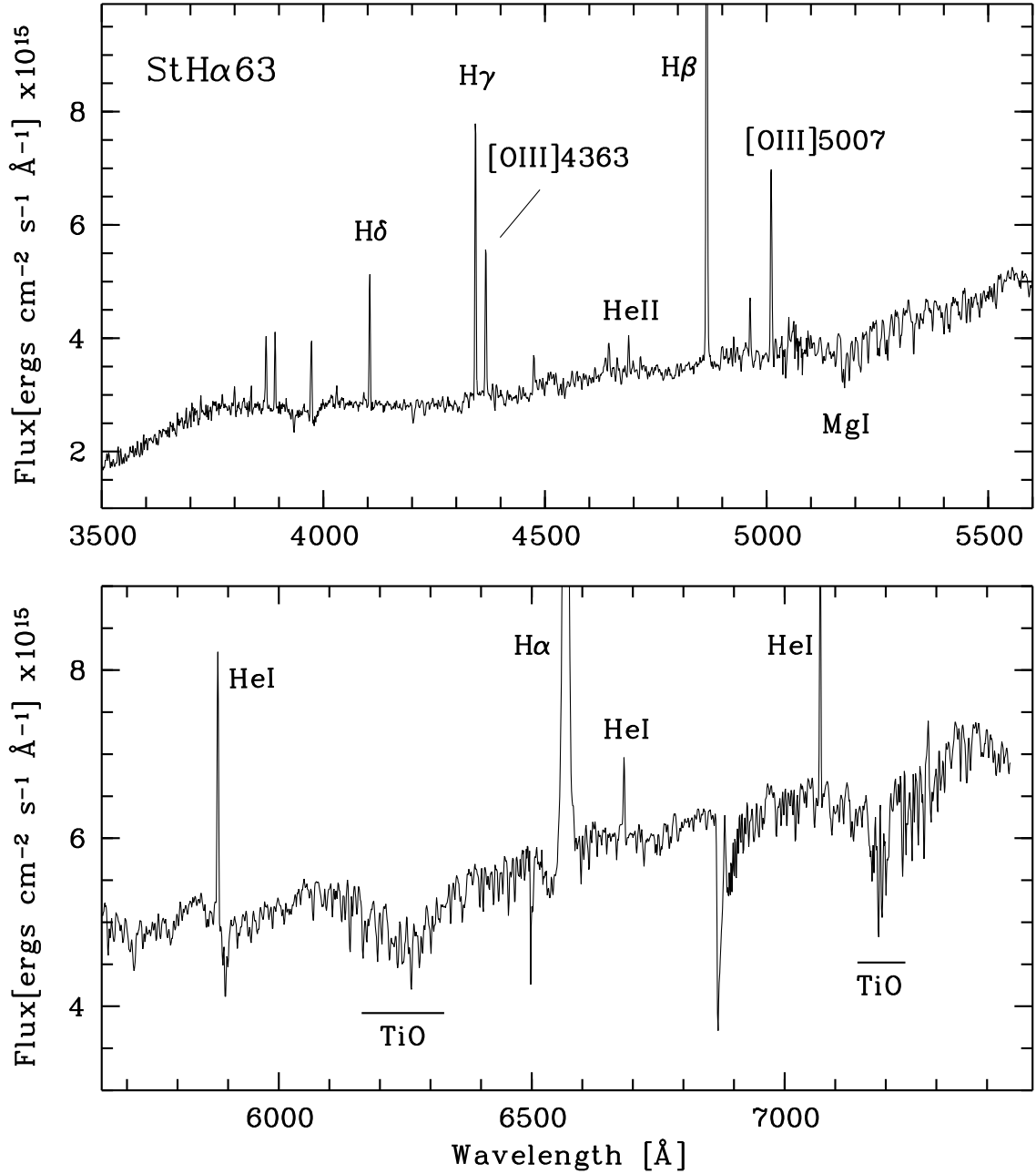}
\caption{Flux calibrated spectra of StH$\alpha$63. Several emission and
  absorption lines, and TiO absorption bands are indicated.}
\end{figure*}

\begin{figure*} 
\includegraphics[angle=0,width=150mm,clip=]{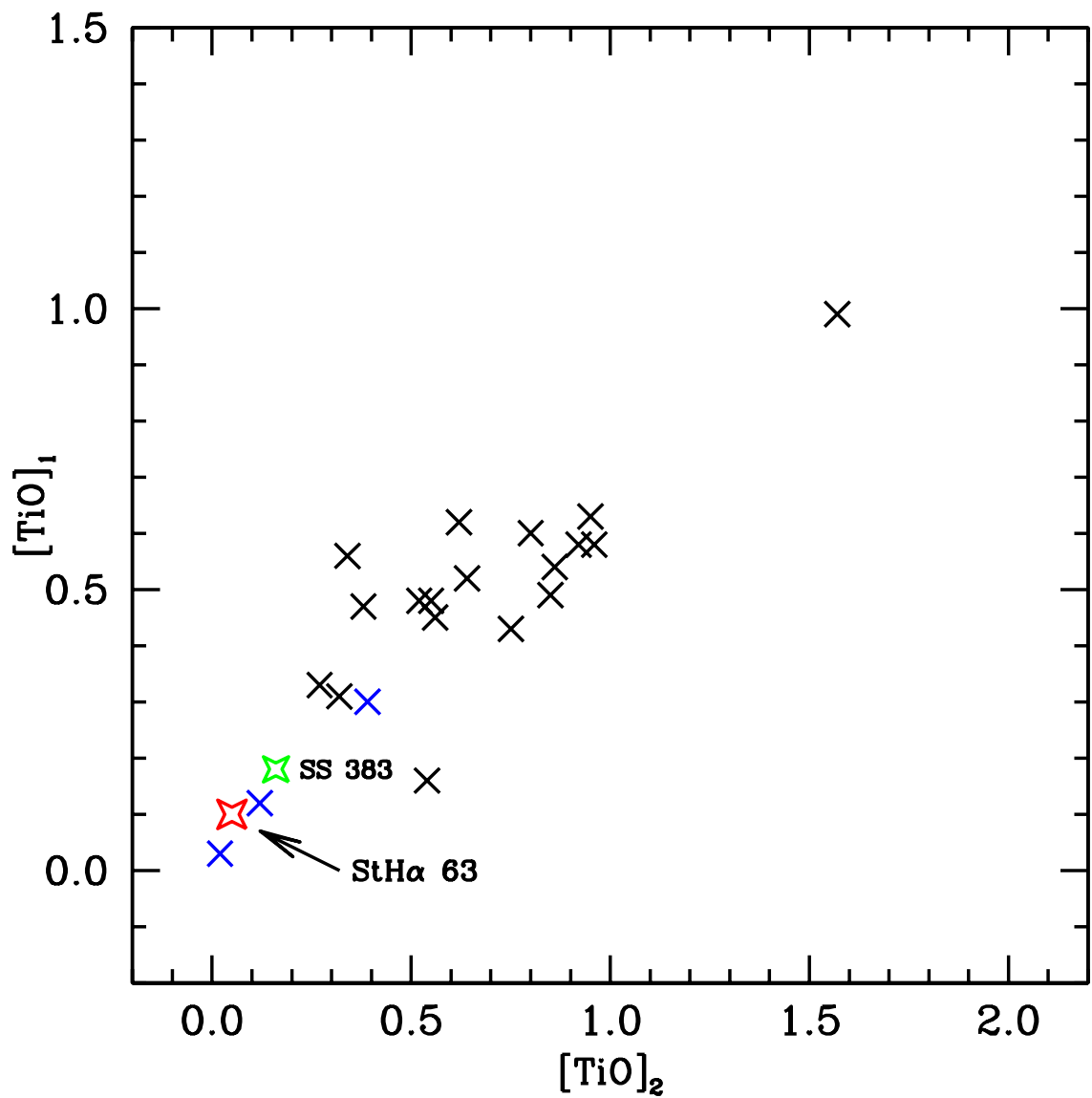}
\caption{Position of StH$\alpha$ 63 ({\sl red star}) in the [TiO]$_{1}$ $vs$
  [TiO]$_{2}$ diagram for a sample of symbiotic stars analyzed by
  Kenyon \& Fern\'andez-Castro (1987b).  Black crosses represent the S-
  and the D-type symbiotics with spectral types later than M0.
  The smaller sample of S-type symbiotics with spectral types not
  later than K9 (the yellow symbiotics) AG Dra, RS Oph and TX CVn
  also analyzed by Kenyon \& Fernandez-Castro (1987b), are represented
  in this diagram by red crosses. The green star represents SS\,383 and the
  red star StH$\alpha$63.}
\end{figure*}

\begin{figure*}
\includegraphics[angle=0,width=140mm,clip=]{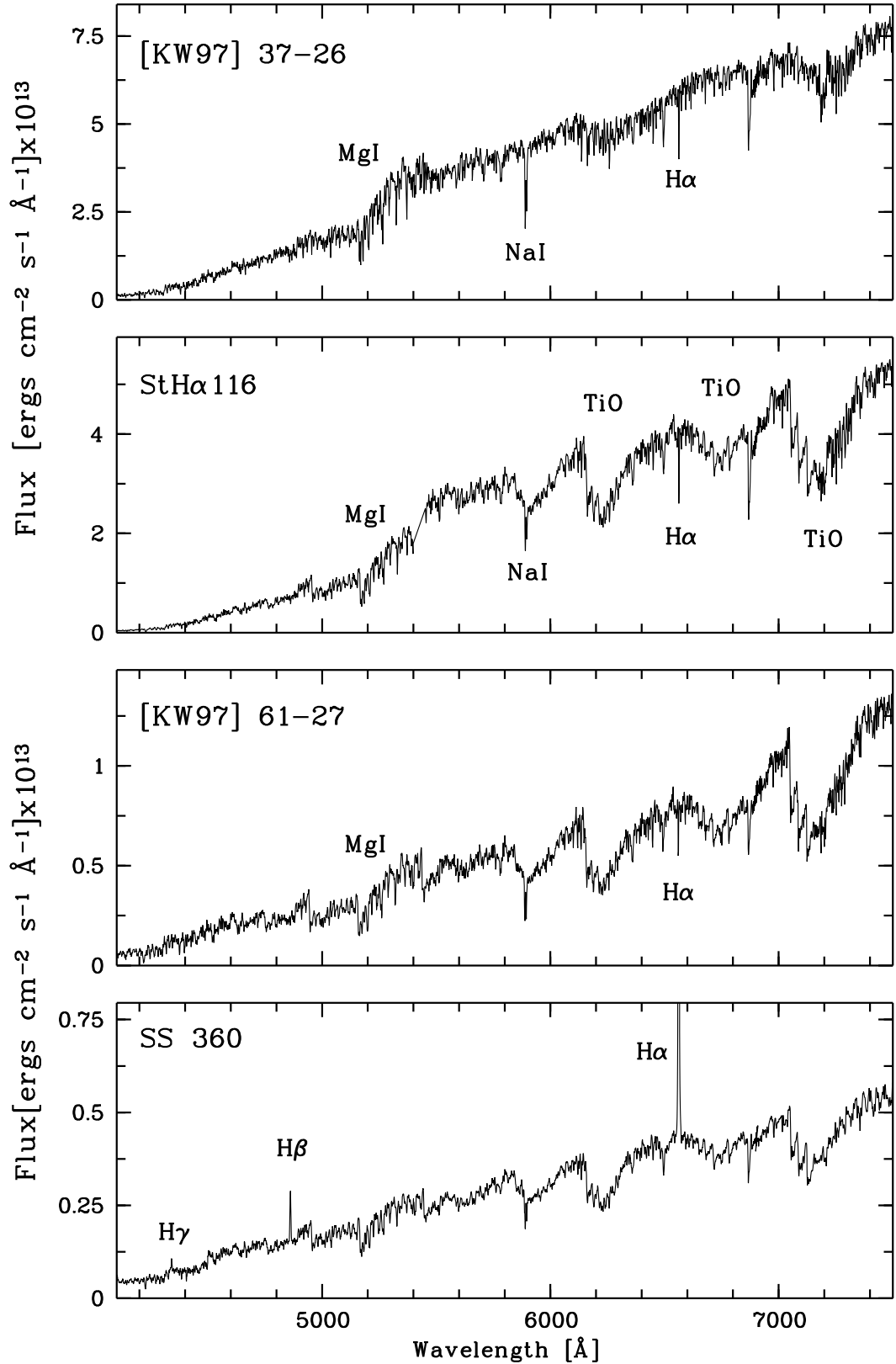}
\caption{Flux calibrated spectrum of the four candidate yellow
  symbiotics [KW97] 37-26, StH$\alpha$116, [KW97] 61-27, and
  SS\,360. Some emission and absorption features are indicated.}
\end{figure*}

\begin{figure*}
\includegraphics[angle=0,width=170mm,clip=]{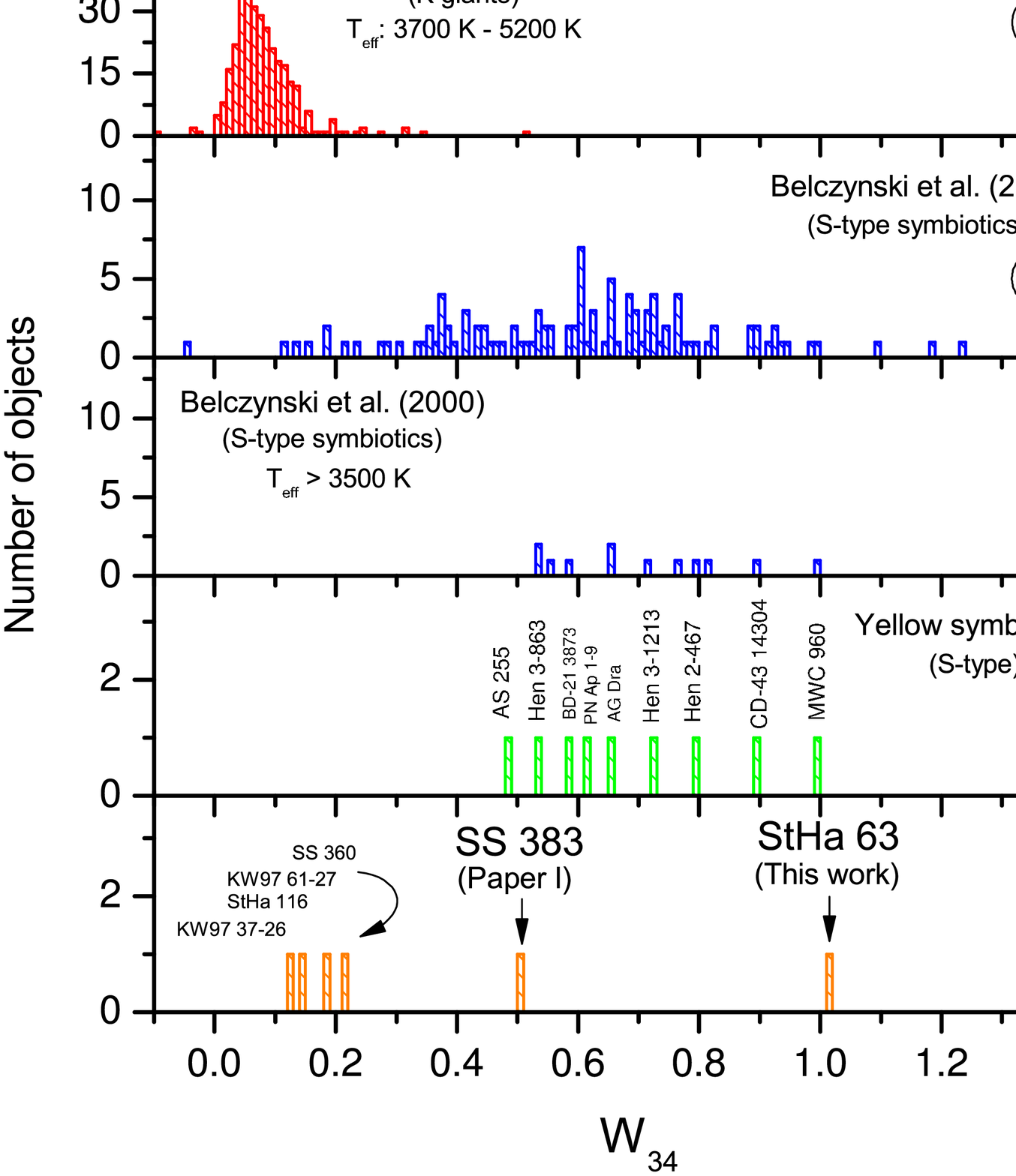}
\caption{Distribution of the WISE W3-W4 (W34) index for five samples
  of K-giant and symbiotic stars. The content of the histograms is
  explained within each of them (see text for details).}
\end{figure*}


\begin{thebibliography}{}


\bibitem{}
Allen, D.A., 1984, PASA, 5, 369.

\bibitem{}
Baella, N.O., 2012, PhD thesis, Observat\'orio Nacional, Rio de Janeiro.

\bibitem{}
Baella, N.O., Pereira, C.B. \& Miranda, L.F., 2013, \aj, 146, 115 (BPM13)

\bibitem{}
Belczy\'nski, K., Mikolajewska, J., Munari, U., Ivison, R.J. \& Friedjung, M., 2000,
Astron. Astrophys. Sup. Ser., 146, 407

\bibitem{}
Bilir, S., Ak, S., Karaali, S., Cabrera-Lavers, A., Chonis, T.S. et al. 2008, \mnras,
384, 1178.

\bibitem{}
Busso, M., Gallino, R., Lambert, D.L., Travaglio, C. \& Smith, V.V. 2001,
ApJ, 557, 802

\bibitem{}
Corradi, R.L.M., Rodr\'{\i}guez-Flores, E.R., Mampaso, A., Greimel, R., Viironen, K. et al. 2008,
\aap, 480, 409

\bibitem{}
Corradi, R.L.M., Valentini, M., Munari, U., Drew, J.E.; Rodr\'{\i}guez-Flores, E.R. et al. 2010,
\aap, 509, 41

\bibitem{}
Cutri, R.M., Skrutski, M.F., van Dyk, S., Beichman, C.A., Carpenter, J.M. et al. 
2003, The IRSA 2MASS All Sky Point Source Catalog, NASA/IPAC Infrared Science 
Archive

\bibitem{}
Downes, R.A. \& Keyes, C.D., 1988, \aj, 96, 777

\bibitem{}
Guti\'errez-Moreno, A. \& Moreno, H., 1996, \pasp, 108, 972

\bibitem{}
Guti\'errez-Moreno, A., Moreno, H. \& Cortes, G., 1995, \pasp, 107, 462

\bibitem{}
Guti\'errez-Moreno, A., Moreno, H. \& Costa, E., 1999, \pasp, 111, 571

\bibitem{}
Hamuy, M., Suntzeff, N.B., Heathcote, S.R., Walker, A.R., Gigoux, P. et al, 1994, \pasp, 106, 566

\bibitem{}
Hekker, S., \& Mel\'endez, J. 2007, \aap, 475, 1003

\bibitem{}
Jacoby, G.H., Hunter, D.A. \& Christian, C.A., 1984, \apjs, 56, 257

\bibitem{}
Jorissen, A., 2003, in  Symbiotic stars probing stellar evolution,
ed.  R.L.M. Corradi, J. Mikolajewska \&  T.J. Mahoney, ASP Conf. Ser., 303, 25

\bibitem{}
Kenyon, S.J.\& Fern\'andez-Castro, T., 1987, \aj, 93, 938

\bibitem{}
Kenyon, S.J., Livio, M., Mikolajewska, J. \& Tout, C.A., 1993, \apj, 407, L81

\bibitem{}
Kingsburgh, R.L. \& Barlow, M.J., 1994, \mnras, 271, 257

\bibitem{}
Kohoutek, L. \& Wehmeyer, R., 1997, Catalogue of stars in the Northern Milky Way having 
H$\alpha$ in emission, Publisher : Hamburg Sternwarte - ESO

\bibitem{}
Lasker, B.M., Lattanzi, M.G., McLean, B.J., Bucciarelli, B., Drimmel, R., et al. 2008,
\apj, 136, 735.

\bibitem{}
Li, J., Mikolajewska, J., Chen, Xue-Fei, Luo, A.-Li, Rebassa-Mansergas, A., 2015c, 
RAA (Research in Astronomy and Astrophysics), 15, 1332

\bibitem{}
MacConnell, D.J., 1981, Astron. Astrophys. Sup. Ser., 44, 387

\bibitem{}
Magrini, L., Corradi, R.L.M. \& Munari, U., 2003, in Symbiotic stars probing stellar
evolution, ASP Conf. Ser., 303, 539

\bibitem{}
Massey, P., Strobel, K., Barnes, J.V. \& Anderson, E., 1988, \apj, 328, 315

\bibitem{}
Maheswar, G., Manoj, P. \& Bhatt, H.C., 2003, \aap, 402, 963

\bibitem{}
Mikolajewska, J. \& Kenyon, S.J., 1992, \aj, 103, 579

\bibitem{}
Mikolajewska, J. 1997, in Mikolajewska J., ed., Physical Processes in Symbiotic
Binaries  and  Related  Objects.  Copernicus  Foundation  for  Polish
Astronomy, Warsaw, p. 3.

\bibitem{}
Miszalski, B., Mikolajewska, J. \& Udalski, A., 2013, \mnras, 432, 3186

\bibitem{}
Miszalski, B. \& Mikolajewska, J., 2014, \mnras, 440, 1410

\bibitem{}
Munari, U. \& Renzini, A., 1992, \apj, 397, L87

\bibitem{}
Munari, U. \& Zwitter, T., 2002, \aap, 383, 188

\bibitem{}
M\"urset, U. \& Schmid, H.M., 1999, Astron. Astrophys. Sup. Ser., 137, 473

\bibitem{}
Osterbrock, D.E. \& Ferland, G.J, 2006, Astrophysics of Gaseous Nebulae and
Active Galactic Nuclei (2nd ed.; Mill Valley, CA: Univ. Science Books)

\bibitem{}
Parker, Q.A., Phillipps, S., Pierce, M.J., Hartley, M., Hambly, N.C., 2005, \mnras, 362, 689

\bibitem{}
Pereira, C.B., Baella, N.O., Daflon, S. \& Miranda, L.F., 2010, \aa, 509, A13

\bibitem{}
Pereira, C.B. \& Miranda, L.F., 2005, \aap, 433, 579

\bibitem{}
Pereira, C.B. \& Miranda, L.F., 2007, \aap, 467, 1249

\bibitem{}
Pereira, C.B. \& Roig, F., 2009, \aj, 137, 118

\bibitem{}
Phillips, J.P., 2007, \mnras, 376, 1120

\bibitem{}
Proga, D., Mikolajewska, J. \& Kenyon, S.J., 1994, \mnras, 268, 213

\bibitem{}
Rodr\'iguez-Flores, E.R., Corradi, R.L.M., Mampaso, A., Garc\'{\i}a-Alvarez, D., Munari, U.
et al, 2014, \aap, 567, 49

\bibitem{}
Schlafly, E.F. \& Finkbeiner, D.P., 2011, \apj, 737, 103.

\bibitem{}
Schlegel, E.M., Finkbeiner, D.P. \& Davies, M., 1998, \apj, 500, 525

\bibitem{}
Schmid, H.M. \& Schild, H., 1990, \mnras, 246, 84

\bibitem{}
Schmid, H.M. \& Nussbaumer, H., 1993, \aap, 268, 159

\bibitem{}
Schwartz, R.D., Persson, S.E. \& Hamann, F.W., 1990, \aj, 100, 793

\bibitem{}
Smith, V.V., Cunha, K., Jorissen, A. \& Boffin, H.M.J., 1996, \aap, 315, 179

\bibitem{}
Smith, V.V., Cunha, K., Jorissen, A. \& Boffin, H.M.J., 1997, \aap, 324, 97

\bibitem{}
Stephenson, C.B. \& Sanduleak, N., 1977, \apjs, 33, 459

\bibitem{}
Stephenson, C.B., 1986, \apj, 300, 779

\bibitem{}
Turnshek, D.E, Turnshek, D.A., Craine, E.R. \& Boeshaar, P.C. 1985, An
Atlas of Digital Spectra of Cool Stars (Tucson: Western Research Company)

\bibitem{}
Wright, E. L. et al. 2010, \aj, 140, 1868

\bibitem{}
Zacharias, N., Monet, D. G., Levine, S. E., et al. 2005, yCat,
1297, 

\end{thebibliography}
\end{document}